# Critical current density of hole doped high-$T_c$ cuprates and heavy fermion superconductors: relevance to the possible quantum critical behavior

S. H. Naqib* and R. S. Islam

Department of Physics, University of Rajshahi, Rajshahi 6205, Bangladesh
*Corresponding author; Email: salehnaqib@yahoo.com

**Abstract**

The superconducting critical current density, $J_c$, in hole doped cuprates show strong dependence on the doped hole content, $p$, within the copper oxides plane(s). The doping dependent $J_c$ mainly exhibits the variation of the intrinsic depairing critical current density as $p$ is varied. $J_c(p)$ tends to peak at $p \sim 0.185$ in copper oxide superconductors. This particular value of the hole content, often termed as the critical hole concentration, has several features putative to a quantum critical point (QCP). Very recently, the pressure dependences of the superconducting transition temperature ($T_c$) and the critical current ($I_c$) in pure CeRhIn$_5$ and Sn doped CeRhIn$_5$ heavy fermion compounds have been reported (Nature Communications (2018) 9:44, DOI: 10.1038/s41467-018-02899-5). The critical pressure demarcates an antiferromagnetic quantum critical point where both $T_c$ and $I_c$ are maximized. We have compared and contrasted this behavior with those found for Y$_{1-x}$Ca$_x$Ba$_2$Cu$_3$O$_{7-\delta}$ in this brief communication. The resemblance of the systematic behavior of the critical current with pressure and hole content between heavy fermion systems and hole doped cuprates is significant. This adds to the circumstantial evidence that quantum critical physics probably plays a notable role beyond the unconventional normal and superconducting state properties of copper oxide superconductors.

## 1 Introduction

It has been well over three decades since the discovery of superconductivity at high transition temperature in hole doped copper oxide materials in the mid-eighties [1, 2]. The precise mechanism leading to Cooper pairing of the extra holes added to the CuO$_2$ planes of these strongly correlated electronic systems remains elusive till date [3 – 5]. A remarkable set of coexisting and often competing electronic ground states [3 – 7] in hole doped cuprates pose a serious theoretical challenge which the condensed matter physics community is yet to surmount. The standard theory for condensed matter physics, the Landau Fermi-liquid theory, breaks down completely in the case of underdoped (UD) cuprates [3 – 5]. The overdoped (OD) side is comparatively more conventional but still exhibits a number of anomalous characteristics [8].

In the absence of any agreed upon theoretical scheme to describe the Mott physics of undoped antiferromagnetic insulating state and its eventual transformation to the pseudogapped normal state, charge and spin density ordered states, and *d*-wave superconductivity upon hole doping, in a coherent fashion, the cuprate research community has focused their attention in exploring various possible scaling relations and generic features found in these materials and in other strongly correlated electronic systems with non-Fermi liquid features [9, 10]. These systematic studies of generic behaviors can provide us with important clues to unlock the mystery of the physics of electronic phase diagram of high-$T_c$ cuprates in the normal and superconducting (SC) states.

Since the early days of cuprate superconductivity, one of the most puzzling features has been the normal state *T*-linear resistivity in these compounds [11 – 14]. In ordinary metals the resistivity saturates at high temperature where the wavelength of the charge carriers becomes comparable to the transport mean free path (Ioffe-Regel limit). More importantly the temperature dependent resistivity, $\rho(T)$, varies as $T^2$ at low temperatures, a canonical Fermi-liquid behavior. Contrary to these normal metallic behaviors, $\rho(T)$ in hole doped cuprates at critical doping shows a completely *T*-linear behavior from near $T_c$ to as high a temperature as can be measured [12]. Besides, the T-dependent Hall coefficient, magnetic excitations, and magnetoresistance in the normal state show non Fermi-liquid-like behavior in these compounds [15 – 17]. Since late 1990s it had been proposed that, presence of a quantum critical point (QCP) in the *T-p* phase diagram could be responsible for unconventional charge and magnetic excitations that could possibly offer explanations for non Fermi-liquid like charge and magnetic transport properties of high $T_c$ cuprates. A QCP is understood via the concept of quantum phase transition (QPT). Unlike ordinary phase transitions driven by thermal energy, a QPT is characterized at a particular value of non-thermal parameter (e.g., critical hole content for cuprates) where a continuous phase transition takes place between a quantum disordered phase and a quantum ordered phase at zero temperature. The correlations at the QCP demonstrate spatio-temporal scale invariance. This implies that the poles present in the quasiparticle (QP) spectral function as predicted in the Fermi-liquid theory are absent here. Instead, we find power-law scaling behavior and the QP spectral function assumes a form given by $\omega/T$. This leads to a dissipative QP relaxation time given by $h/(2\pi k_B T)$ which in turn implies that the scattering rate is linear in *T*. Moving away from the QCP, the energy scale for which scale invariance is valid gradually increases [18]. As a consequence, in the *T-p* plane (considering electronic phase diagram of cuprates) there is a quantum critical wedge opening up from the QCP. This feature is seen in the generic *T-p* phase diagram of all hole doped cuprates [13, 18].

Quite interestingly the singular interactions arising from the competing phases at the QCP can provide with the 'glue' for Cooper pairing at high temperatures [19 – 21]. For example, Castellani et al. [19] presented a scenario where a QCP due to formation of incommensurate charge density waves roughly accounts for some of the generic features of the high-$T_c$ cuprates, both in the normal and in the SC states. Specifically, the singular interaction arising close to this charge-driven QCP gives rise to non-Fermi liquid behavior universally found at the vicinity of the optimal doping. This interaction can also lead to *d*-wave Cooper pair formation with a $T_c$ strongly dependent on the hole content.

It should be mentioned that the presence and precise nature of a QCP in hole doped cuprates are hotly debated issues [22]. The situation is clearer in the case of heavy fermion (HF) and iron pnictide superconductors [22 – 24]. It was not until the last two decades unambiguous signs of QCP had been experimentally observed in the HF systems. For example, in $CeCu_{6-y}Au_y$, Au-doping turns the paramagnetic metal phase of the pure $CeCu_6$ into an antiferromagnetic (AF) metal phase for $y > 0.1$. As a function of the Au concentration, the Neel temperature, $T_N$ sets in continuously; the same is the case for the low-temperature staggered moment – the spatially-modulated magnetization at the AF wave vector [22, 23]. These characteristics establish the existence of an AF QCP quite clearly. It also induces *T*-linearity in the temperature dependent resistivity [23]. As far as SC HFs are concerned, $CeCu_2Si_2$ is the prime candidate for antiparamagnon mediated superconductivity near a spin density wave QCP [24]. Besides, superconductivity may emerge from the proximity to a magnetic field-induced QCP, like that in $CeCoIn_5$ [25] and perhaps also in $UBe_{13}$ [26]. The case for QCP in iron pnictides is quite strong. Evidence for superconductivity in at least one iron pnictide due to quantum critical spin fluctuation is overwhelming [22, 27]. Identification of the ground electronic state, its symmetry and thermodynamic signature of the symmetry breaking at the QCP in cuprates, on the other hand, are unclear [22] at the moment. Under the circumstances, a useful strategy is to compare and contrast various non-Fermi liquid like properties of cuprates with those of the HF and iron pnictide systems. *T*-linear resistivity at the critical doping (or other relevant non-thermal parameters like pressure and magnetic field at the QCP) has been the most widely studied phenomenon. Very recently, Jung et al. [28] have studied the SC critical currents, $I_c$, in $CeRhIn_5$ and 4.4% Sn-doped $CeRhIn_5$ HF superconductors as a function of pressure (*P*). The $I_{c0}$s (zero-field critical currents) of these HF compounds under pressure exhibit a universal temperature dependence, underlining that the peak in zero-field $I_{c0}(P)$ is determined predominantly by quantum critical fluctuations associated with a hidden magnetic QCP at a critical pressure $P_c$, where superconducting transition temperature is also maximum. Motivated by this particular study [28], we have investigated the hole content dependent zero-field critical current density, $J_{c0}$, of a series of $Y_{1-x}Ca_xBa_2Cu_3O_{7-\delta}$ superconductors over wide range of compositions. We have also looked at the hole content dependent vortex activation energy and irreversibility field of $YBa_2CuO_{7-\delta}$ thin films in this investigation. The generic behaviors of the superconducting critical current density and vortex pinning characteristics in $Y_{1-x}Ca_xBa_2Cu_3O_{7-\delta}$ and Ce-based HF superconductors show strikingly similar behavior. We have discussed this feature and their possible implication in this short communication. This is the first comparative systematic study based on critical current density between hole doped cuprates and heavy fermion superconductors to the best of our knowledge.

The rest of the paper is organized as follows. A brief description of $Y_{1-x}Ca_xBa_2Cu_3O_{7-\delta}$ compounds and some details regarding the previous $J_c$ and magnetic field dependent resistivity measurements are presented in Section 2. The results are presented and compared to those obtained for Ce-based HF superconductors in Section 3. Section 4 comprises of the discussion on the results and important conclusions of this study.

## 2 $Y_{1-x}Ca_xBa_2Cu_3O_{7-\delta}$ samples and measurements

High-quality c-axis oriented crystalline thin films of $Y_{1-x}Ca_xBa_2Cu_3O_{7-\delta}$ (x = 0.00, 0.05, 0.10, 0.20) were grown on $SrTiO_3$ substrates using the method of pulsed LASER ablation technique. Substrates of dimensions 5 x 5 x 1 mm$^3$ and 10 x 5 x 1 mm$^3$ were used. The thicknesses of the films lie within 2800 ± 300 Å. Details regarding the film preparation and characterization can be found in Ref. [29]. Hole content within the $CuO_2$ planes were varied by two independent means. The oxygen deficiency, $\delta$, in the $CuO_{1-\delta}$ chains were controlled via oxygen annealing at different temperatures and partial pressures. The Ca content, x, substituted for the Y atom in the charge reservoir layer also adds holes to the $CuO_2$ planes independent of the oxygen loading in the $CuO_{1-\delta}$ chains. This enables one to access the overdoped side relatively easily. Pure YBCO with fully oxygen loaded CuO chains can give a maximum p value ~ 0.180. Information about the annealing treatments and magnetization measurements of the thin films can be found in Refs. [29 – 31]. The hole content was estimated with high degree of accuracy from three different methods: room temperature thermopower ($S$[290 K]) [32, 33], c-axis lattice constant [30], and the well known parabolic $T_c(p)$ relation [34, 35]. In this paper we have used the p-values obtained from the $S$[290 K] data. This is quite insensitive to the crystalline order and disorder content of the sample and depends solely on the number of doped holes in the $CuO_2$ plane. Details regarding the magnetic field dependent resistivity ($\rho_{ab}(H, T)$) measurements and analysis of the flux dynamics can be found in Ref. [36]. All the measurements presented for $Y_{1-x}Ca_xBa_2Cu_3O_{7-\delta}$ in this study were done for the *H* ΙΙ *c* configuration, where the supercurrent circulated in the $CuO_2$ plane. We have shown representative M-H loops for $Y_{1-x}Ca_xBa_2Cu_3O_{7-\delta}$ thin films at different temperatures and hole contents in Figs. 1. Representative $\rho_{ab}(H, T)$ data for $YBa_2Cu_3O_{7-\delta}$ thin films are shown in Figs. 2.

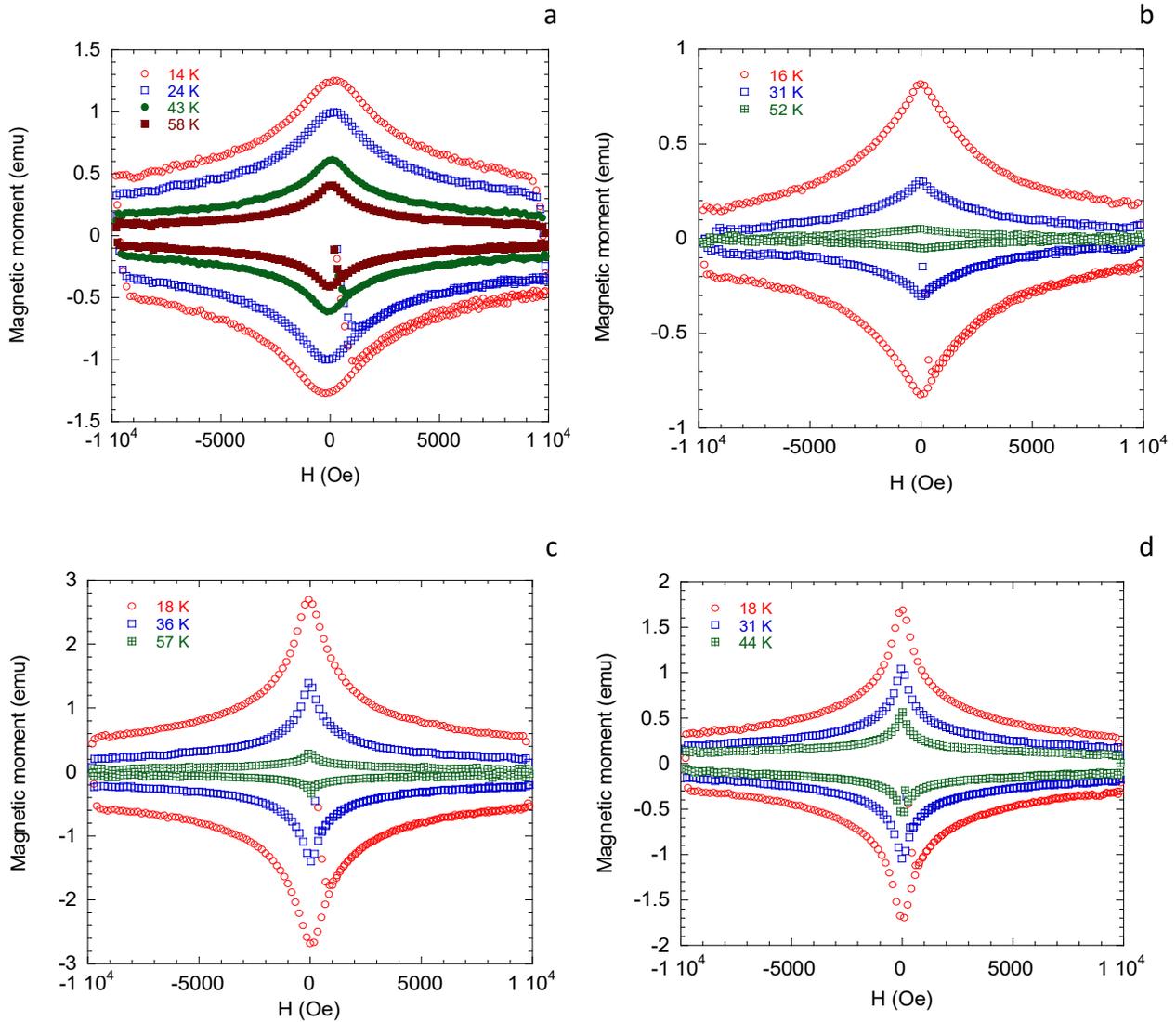

Figure 1: Representative magnetization loops for $Y_{1-x}Ca_xBa_2Cu_3O_{7-\delta}$ thin films with different compositions at different temperatures. The magnetic field was applied along *c*-direction. The sample compositions and hole contents are (a) $YBa_2Cu_3O_{7-\delta}$; *p* = 0.162, (b) $Y_{0.95}Ca_{0.05}Ba_2Cu_3O_{7-\delta}$; p = 0.123, (c) $Y_{0.90}Ca_{0.10}Ba_2Cu_3O_{7-\delta}$; p = 0.198, and (d) $Y_{0.80}Ca_{0.20}Ba_2Cu_3O_{7-\delta}$; p = 0.144. *p*-values are accurate within ± 0.004. For clarity only one in twenty data points are shown.

The zero-field critical current, $J_{c0}$, for the $Y_{1-x}Ca_xBa_2Cu_3O_{7-\delta}$ thin films with different amounts of Ca and oxygen deficiencies were calculated from the width of the magnetization loops at $H = 0$ G and the dimensions of the thin films following the method developed by Brandt and Indenbom [37] for finite geometry with the modified critical state model.

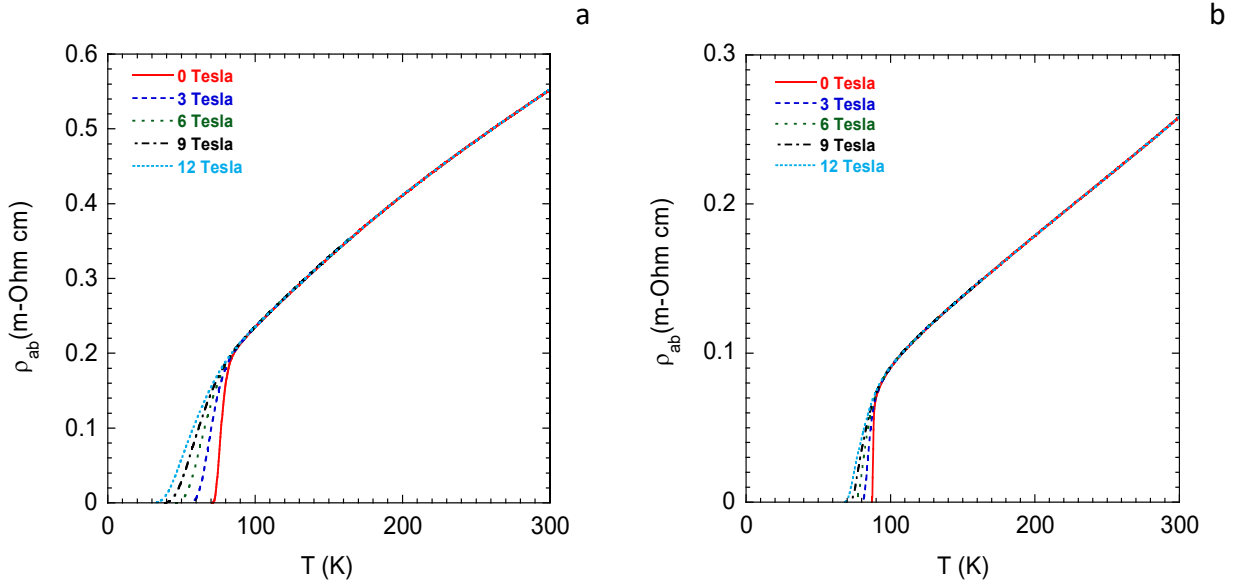

Figure 2: Representative magnetic field-dependent in-plane resistivity data for $YBa_2Cu_3O_{7-\delta}$ thin films with different hole contents. The magnetic fields were applied along the *c*-direction. The hole contents are (a) 0.118 and (b) 0.170. These values are accurate within ± 0.004.

**3 Hole content dependent critical current density**

The critical current density depends strongly on temperature. In this study, we have used the zero temperature critical current density for comparison. This was done by fitting the hole content dependent zero-field critical current density to the following relation [36, 38]

$$J_{c0}(\tau) = J_0(1-\tau)^n \tag{1}$$

where, $\tau = (T/T_c)$, is the reduced temperature and $J_0$ is the extrapolated zero-field critical current density at $T = 0$ K. Value of the exponent depends on the details of flux pinning properties [38]. For the sample compositions used in this study, the values of *n* lie within the range 2.00 ± 0.60. The value of the exponent, *n* increases systematically with underdoping. The extracted values of $J_0$ for different hole contents are shown in Table 1.

Table 1

Zero-field and zero-temperature critical current density of $Y_{1-x}Ca_xBa_2Cu_3O_{7-\delta}$ thin films.

| Compound | Hole content (p) | Critical current density, $J_0$ ($10^6$ A/cm$^2$) | Normalized critical current density |
|---|---|---|---|
| $YBa_2Cu_3O_{7-\delta}$ | 0.162 | 20.62 | 0.668* |
| | 0.146 | 14.84 | 0.481* |
| | 0.102 | 5.99 | 0.194* |
| | | | |
| $Y_{0.95}Ca_{0.05}Ba_2Cu_3O_{7-\delta}$ | 0.184 | 30.88 | 1.000 |
| | 0.170 | 26.04 | 0.843 |
| | 0.156 | 22.10 | 0.716 |
| | 0.123 | 12.10 | 0.392 |
| | | | |
| $Y_{0.90}Ca_{0.10}Ba_2Cu_3O_{7-\delta}$ | 0.198 | 23.73 | 0.831 |
| | 0.188 | 28.54 | 1.000 |
| | 0.162 | 23.50 | 0.823 |
| | 0.160 | 24.41 | 0.855 |
| | 0.126 | 13.08 | 0.458 |
| | | | |
| $Y_{0.80}Ca_{0.20}Ba_2Cu_3O_{7-\delta}$ | 0.201 | 17.08 | 0.921 |
| | 0.186 | 18.54 | 1.000 |
| | 0.166 | 17.01 | 0.917 |
| | 0.150 | 12.98 | 0.700 |
| | 0.144 | 12.03 | 0.649 |
| | 0.136 | 10.09 | 0.544 |

*Normalized with $J_0 = 30.88 \times 10^6$ A/cm$^2$. See Section 4 for details.

We have plotted the normalized zero-temperature zero-field critical current density for $Y_{1-x}Ca_xBa_2Cu_3O_{7-\delta}$ thin films in Fig. 3. $J_0(p)$ has been normalized with the maximum value of $J_0$ for each Ca content (x). It is important to note that irrespective of the Ca content and oxygen deficiency in the $CuO_{1-\delta}$ chain, $J_0(p)$ is maximized when $p \sim 0.185$. We have also shown the normalized zero-field critical current for the 4.4% Sn-doped $CeRhIn_5$ HF superconductor as a function of pressure (P) in the inset. The systematic behavior of doped high-$T_c$ cuprates and the HF compounds as functions of hole content and pressure are strikingly similar, as far as the critical current is concerned.

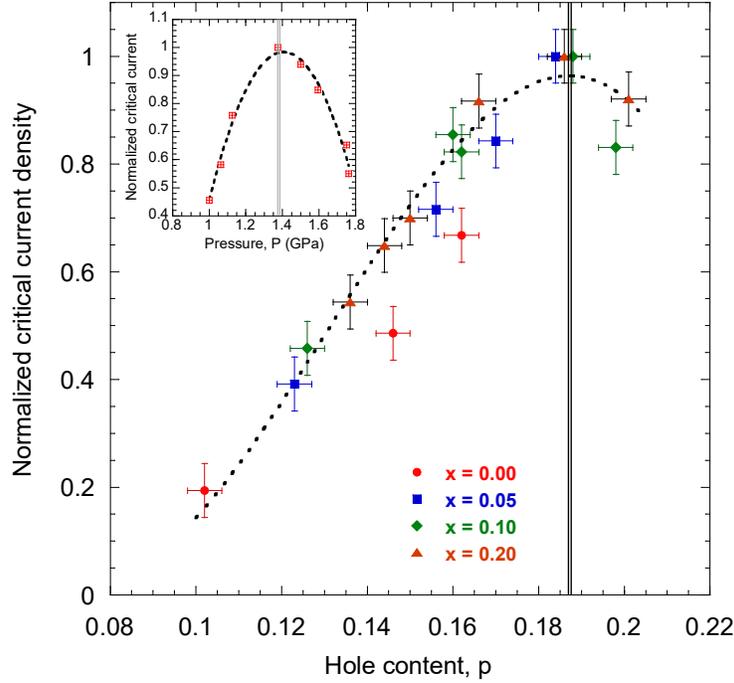

Figure 3: (Main) The normalized zero-temperature and zero-field critical current density of $Y_{1-x}Ca_xBa_2Cu_3O_{7-\delta}$ thin films as a function of doped hole content in the $CuO_2$ planes. The inset shows the variation of the normalized zero-field critical current of 4.4% Sn-doped $CeRhIn_5$ HF superconductor with pressure. The vertical lines mark the maximum critical currents and give the critical hole concentration and the critical pressure, at the putative quantum critical point.

Next, we have shown the $p$-dependent behavior of the characteristic magnetic field, $H_0$, for $YBa_2Cu_3O_{7-\delta}$ thin films in Fig. 4. $H_0(p)$ gives a direct measure of the vortex activation energy and the irreversibility magnetic field [36, 38 – 41]. Resistive broadening of the superconducting transition region as seen in Fig. 2, can be analyzed using the thermally assisted flux flow (TAFF) model [36, 41]. The vortex activation energy, or equivalently the vortex pinning energy, $U(H, T)$, can be expressed quite well in a dimensionless form as follows: $U(T, H) = (1-t)^m (H_0/H)^{-\beta}$. Here, $t = T/T_c$, the reduced temperature, $\beta$ is a constant close to unity, and $m$ is an exponent which varies with hole content, anisotropy and nature of the pinning centers within the sample. From the analysis of the $\rho_{ab}(H, T)$ data for $YBa_2Cu_3O_{7-\delta}$ thin films with different hole concentrations, $H_0(p)$ was calculated [36].

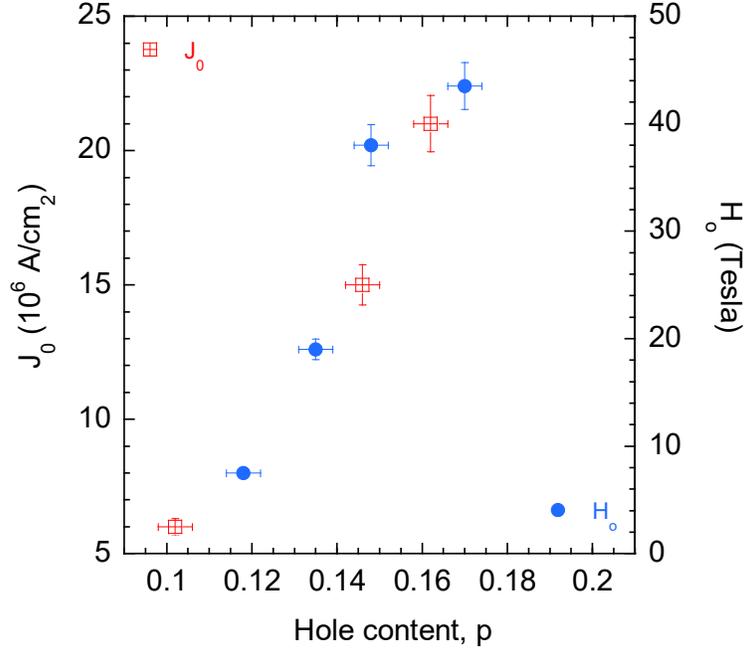

Figure 4: $J_0(p)$ and $H_0(p)$ of $YBa_2Cu_3O_{7-\delta}$ thin films.

It is worth noting that both $J_0(p)$ and $H_0(p)$ changes with the doped hole content in the same fashion in $YBa_2Cu_3O_{7-\delta}$. Therefore, it is reasonable to assume that the $p$-dependent zero-field and zero-temperature critical current density in $Y_{1-x}Ca_xBa_2Cu_3O_{7-\delta}$ actually reflects the doping dependent vortex activation energy, which in turn is closely linked to the $p$-dependent condensation energy and superfluid density of the Cooper pairs [36, 42].

**4 Discussion and conclusions**

The normalized critical current density as a function of in-plane doped hole content of $Y_{1-x}Ca_xBa_2Cu_3O_{7-\delta}$ shows strong resemblance to the normalized critical current of $CeRhIn_5$ and 4.4% Sn-doped $CeRhIn_5$ HF superconductors as a function of pressure. For the high-$T_c$ cuprate, critical current density peaks at $p_c \sim 0.185$, whereas for the 4.4% Sn-doped $CeRhIn_5$ HF compound the critical current peaks for $P_c \sim 1.35$ GPa. These particular values of the control parameters are known as the critical hole concentration and critical pressure, respectively. The importance of critical hole concentration in cuprates has been described in several earlier studies [13, 21, 22, 43, 44] in details. The widely investigated pseudogap in the quasiparticle energy spectrum tends to vanish abruptly at this particular hole content [13, 21, 43, 44], the superfluid density and the superconducting condensation energy is maximized [21, 43], the Fermi surface goes through a reconstruction [45], quasiparticle peaks appear abruptly in the normal state ARPES spectra [43], among others.

Quantum criticality describes the collective excitations in strongly correlated systems undergoing a second-order phase transition at zero temperature. How these excitations can lead to formation of Cooper pairs is a matter of intense interest [46, 47]. There are strong empirical evidences that spin density wave type quantum criticality can lead to superconductivity [22, 24 – 26] but a coherent theoretical scheme is yet to be developed. In recent years a number of attempts have been made to formulate quantum critical SC theory to describe the high $T_c$ and the non-FL behavior of copper oxide superconductors. Wang and Chubukov [20] have considered spin-mediated superconducting pairing at the antiferromagnetic QCP with an ordering momentum of $2k_F$ ($k_F$ is the Fermi momentum). Kivelson et al. [48] studied the effect of soft critical collective fluctuations at a nematic quantum critical point on superconductivity. It was found that Cooper pairing channel is strengthened by such collective modes. Very recently Abanov et al. [49] considered a quantum-critical metal with interaction mediated by fluctuations of a critical order parameter. This interaction gives rise to two competing tendencies – Cooper pairing and non-Fermi liquid behavior, and seems to reproduce a number of anomalous features seen in the electronic phase diagram of hole doped cuprates. It is important to note that, irrespective of the details and the precise nature of the QCP [19, 20, 48, 49], all these proposed models predict enhanced superconductivity at the QCP and therefore, provides us with scenarios where the intrinsic critical current density is maximized at the QCP due to its dependence on the SC condensation energy and superfluid density.

It is worth noticing that in variety of SC systems [22, 50] other than the hole doped cuprates, the quantum critical point coincides with the particular value of control parameter where the SC critical temperature is maximized. In most hole doped cuprates the optimum hole content, $p_{opt} \sim 0.16$, differs from the critical hole concentration, $p_c \sim 0.19$. This probably implies that one parameter scaling of quantum critical behavior is probably not adequate [51] in hole doped cuprates and a separate critical component competing with superconductivity may exist at $p_c \sim 0.19$.

It is not surprising to find that $J_0$ and the characteristic magnetic field $H_0$ follow the same $p$-dependence for YBa$_2$Cu$_3$O$_{7-\delta}$ thin films since $H_0$ gives a measure of the vortex activation energy. We predict the similar pressure dependent behavior of the critical current density and vortex pinning energy for pure and doped CeRhIn$_5$ HF superconductors.

It should be noted that we have used the maximum value of the 5% Ca substituted compound to normalize the critical current density of the Ca-free thin film. This is done because these two films show almost similar physical properties including the residual resistivity, slope of the temperature dependent resistivity and SC transition temperature. For example, the maximum $T_c$ at the optimum hole content ($p$ = 0.16) for $Y_{1-x}Ca_xBa_2Cu_3O_{7-\delta}$ and $YBa_2Cu_3O_{7-\delta}$ thin films are 92.5 K and 91.0 K, respectively. This possibly introduces a small systematic error in the normalized critical current density of $YBa_2Cu_3O_{7-\delta}$. This error has no significant bearing on the conclusions drawn in this paper.

**Acknowledgements**

The authors would like to thank Dr. Anita Semwal for her part in the magnetization measurements.